\documentclass[a4paper,twocolumn,aps,prr,superscriptaddress,showpacs]{revtex4-1}
\usepackage{amsmath,amssymb,bm,graphicx,xcolor}
\usepackage[colorlinks=true,allcolors=blue]{hyperref}
\usepackage[displaymath,mathlines]{lineno}
\usepackage{wrapfig, blindtext}
\usepackage{enumerate}

\begin{document}

\title{Model for creep failure with healing}
\author{Subhadeep Roy}
\email{subhadeeproy03@gmail.com}
\affiliation{PoreLab, Department of Physics, Norwegian University of Science and Technology, NO7491 Trondheim, Norway.}
\author{Takahiro Hatano}
\email{hatano@ess.sci.osaka-u.ac.jp}
\affiliation{Department of Earth and Space Science, Osaka University, 560-0043 Osaka, Japan.}
\date{\today}

\begin{abstract}
To understand general properties of creep failure with healing effects, we study a mean-field fiber bundle model with probabilistic rupture and rejoining processes. The dynamics of the model is determined by two factors: bond breaking and formation of new bonds. Steady states are realized due to the balance between breaking and healing beyond a critical healing factor, below which the bundle breaks completely. Correlation between the fluctuating value of strain generated in the model with time at the steady state leads to a characteristic time that diverges in a scale-free manner as we approach the critical healing factor. Transient behaviors in strain rate also involve a power law with a non-universal exponent. 
\end{abstract}


\maketitle

\section{Introduction}
Fracture phenomena in disordered systems constitute an important problem in various contexts such as industries, mechanical engineering, materials science, earth and planetary science, and statistical physics \cite{Chak1,Herrmann}. Historically, statistical physics has aimed at the competition  between two kinds of randomness: thermal fluctuation and structural disorder. This problem is also important in fracture phenomena: how the failure dynamics is affected by the structural randomness and temperature.

Generally, rupture dynamics is sensitive to temperature. At low temperatures, materials fail immediately if the applied stress reaches the critical value \cite{Duba2013,Wonga2012}. This is referred to as brittle failure. In contrast, enduring slow deformation precedes rupture at moderate or high temperatures. The latter failure mode is referred to as creep failure \cite{Nechad2005,Nechad12005}. Note that creep failure occurs at the stress lower than the critical value. 

Due to its enduring nature, creep failure may be accompanied by healing. This is unlikely for open cracks under tensile stress condition, but more likely for shear failure, where cracks are still in contact after failure. Since healing processes can be also sensitive to temperature, failure dynamics with a healing process provides us with an intriguing problem from the viewpoint of statistical physics. 

A phenomenon in which healing plays a vital role is friction, where the protrusions of two surfaces are jointed. Friction force is supported by these jointed protrusions, which are referred to as asperities \cite{Rabinowicz}. The asperities are detached and rejoined repeatedly by the slip motion, which may be regarded as failure and healing, respectively.
Another interesting phenomenon is the shear failure of snow:  i.e., avalanches. In general, snow is a highly disordered porous material that can heal, and thus exhibits a rich response to external stress \cite{Schneebeli04}.
 Although these phenomena are so complex and involve various elementary physical processes, simple model studies may be useful for extracting and understanding their essential mechanisms.

In this paper, we consider the effect of healing on the dynamics of shear deformation and the subsequent failure by proposing a simple stochastic model. The model we have adopted here is a variation of the fiber bundle model (FBM) \cite{Pierce,Daniels,Fiber1}. Due to its simplicity, the model allows various kinds of extension and thus has been widely utilized to analyze the failure of disordered materials. For instance, although the original FBM does not explicitly include the effect of temperature, there have been some attempts to extend the model to study thermally induced failures \cite{Pradhan07,Yoshioka10,Yoshioka12,royhatano1,royhatano2} or noise-induced failures \cite{Lawn,Coleman,Scorretti,Roux,Pradhan03}, in which the effect of temperature is naturally incorporated.

As opposed to the conventional model, these extended models with finite temperature evolve at a constant external load, and therefore exhibit rich dynamic behaviors. In particular, a model for snow avalanche has been proposed making use of the FBM \cite{Reiweger}. Along the line of these studies, we further extend the FBM to include the healing process. We discuss the time evolution of the model as well as its statistical nature of fluctuations.

In the next section, we explain the conventional fiber bundle model, the algorithms for probabilistic rupture and healing. In section III, we present the numerical results followed by discussions on the work and the concluding remarks in section IV. 


\section{Description of the model}\label{secII}
\subsection{Outline}
The model investigated here is analogous to the equal-load sharing fiber bundle model (ELS-FBM), but here we have made some modifications. Below we explain the model, starting with the features common to the conventional one.

Initially, the model consists of $N$ Hookean fibers connected in between two parallel rigid bars, at which the external force $F$ is applied. The force $F$ is redistributed among the constituent fibers. In the equal-load sharing model, which the present study involves, the force $F$ is equally distributed to all the remaining fibers. Namely, the force on each fiber is $F/N$, which is hereafter referred to as the initial stress, $\sigma_0 := F/N$. Note that this is one of the important parameters in this model.

During time evolution, some fibers may break and thus cannot support the load. Writing the number of intact fibers at time $t$ as $L_t$, hereafter we use the intact fraction $\phi_t$ defined by $\phi_t := L_t/N$. The stress at time $t$, denoted by $\sigma(t)$, is then equal to $\sigma_0/\phi_t$. 

\subsection{Fiber strength}
The fibers may break under the stress. Each fiber has its own failure strength, which is drawn randomly from the distribution function, $\rho(h)$. Although it is difficult to determine the distribution function in experiments, the fracture strength of micro-elements may not be distributed over several orders of magnitude. We may thus approximate such distribution using a uniform distribution with the mean of $\bar\sigma$ and the half-width of $\delta$:
\begin{equation}\label{eq:threshold}
\rho(h) = \begin{cases}
    \displaystyle\frac{1}{2\delta},  & (\bar\sigma-\delta \le h \le \bar\sigma+\delta) \\
    0 . & ({\rm otherwise})
  \end{cases}
\end{equation}
The half-width of distribution, $\delta$, is the measure of disorder in this model.

\subsection{Probabilistic failure}
In the conventional model, any fiber breaks once the stress exceeds its threshold value. This process is deterministic. In contrast, in the present model, the rupture criterion of each fiber is probabilistic. We introduce the following function for the rupture probability per unit time for fiber $i$ at time $t$.
\begin{equation}\label{eq:prob_failure}
P_r (t,i) = k_1e^{-E(t,i)/k_B T}
\end{equation}
where $k_1$ is the rate constant and $E(i,t)$ is an activation energy for the rupture of fiber $i$ at time $t$. This is written using the shear modulus $G$ and the volume constant $\Omega$ as
\begin{equation}\label{eq:energy}
E (t,i) = \displaystyle\frac{\Omega}{2G}[h(i)^2-\sigma (t)^2],
\end{equation}
where $h(i)$ and $\sigma (t)$ are, respectively, the strength of fiber $i$ and the stress at time $t$. If $h(i) \le \sigma (t) $, fiber $i$ breaks with probability $1$ irrespective of Eq. (\ref{eq:prob_failure}).

\subsection{Probabilistic healing}
Another important difference from the conventional model is the healing process. Namely, broken fibers can rejoin with a certain probability. During unit time, a broken fiber can heal up at the probability of $P_h(t)$.
\begin{equation}\label{eq:prob_healing_0}
P_{h}(t) = k_2(1-\phi_t)e^{-A/k_BT},
\end{equation} 
where $k_2$ is another rate constant, $A$ is an activation energy for healing, and $\phi_t$ is the fraction of intact fibers at time $t$: $\phi_t = L_t/N$.

The factor of $(1-\phi_t)$ in Eq. (\ref{eq:prob_healing_0}) means that the healing probability of a single broken fiber is proportional to the number of broken fibers. This implies that healing can occur when a broken fiber meets another broken one. In this respect, the rate constant $k_2$ may be interpreted as the collision frequency of broken fibers. This further suggests that $k_2$ may be proportional to the slip velocity or shear rate, since the collision frequency of fibers is generally proportional to the slip velocity for shear deformation. The healing probability is large when the shear rate is large, since there are more chances for a broken fiber to be in contact with other broken ones. In the context of friction, this is the frequency at which a protrusion meets another. It is thus proportional to the slip velocity. 

We can rewrite the healing probability in a simpler manner.
\begin{equation}\label{eq:prob_healing}
P_{h}(t) = \omega (1-\phi_t),
\end{equation} 
with $\omega := k_2 \exp(-A/k_BT)$. The independent parameters are then reduced from $(k_2,A)$ to $\omega$.
The intrinsic rate constants are thus $(k_1, \omega)$. Note that $\omega$ depends on the temperature, while $k_1$ does not.

A rejoined fiber is assigned a new strength (threshold stress) according to the distribution function given by Eq. (\ref{eq:threshold}). We assume that a certain fiber is allowed to rejoin infinite times.

\subsection{Time evolution}
Although the conventional FBM does not have the concept of time, there have been some attempts to extend the model to describe time evolution \cite{Pradhan07,royhatano1}. The basic idea is to introduce the duration required for the stress redistribution. More precisely, the rupture of a single fiber and the subsequent stress redistribution may take a certain time. This duration is denoted by $\tau$, which is an intrinsic time scale in the model.
The time evolution of the present system thus consists of the accumulation of this intrinsic time scale: $t \rightarrow t+\tau$. Namely, $\tau$ constitutes a single time step. 
Since $\tau$ is an intrinsic time constant, the time evolution here is inherently discrete and we cannot consider the continuous limit: $\tau\rightarrow 0$.

During this single time step, the rupture probability for an intact fiber is $\tau P_r (t)$, and the healing probability for a broken fiber is $\tau P_h(t)$.

At $t=0$, the force on each fiber is $\sigma_0$. The failure probability per unit time $P_r(0,i)$ is then computed for each fiber using Eqs. (\ref{eq:prob_failure}) and (\ref{eq:energy}). A random number $R(0,i)$ is drawn from the uniform distribution defined on the interval $[0,1]$. If $R(0,i) < \tau P_r(0,i) $, fiber $i$ breaks and is removed in the next time step ($t=\tau$). This process is repeated for all the fibers.

In the subsequent time steps ($t \ge \tau$), all the fibers are inspected in order: $i=1, 2, \cdots, N$. If fiber $i$ is intact, the above procedure is repeated to check if it fails or not. If fiber $i$ is broken, we check if it heals or not. This is done by drawing another random number $R^{\ast}(t,i)$ from the uniform distribution on $[0,1]$. If $R^{\ast}(t,i) < \tau P_h(t)$, fiber $i$ heals up and a new threshold is assigned to it chosen from the distribution given by Eq. \ref{eq:threshold}. 

After inspecting all the fibers in this manner, the number of surviving fibers $L_t$ is updated to $L_{t+\tau}$ and so is the force per fiber.

\subsection{Dimensionless quantities}
As discussed above, in this model, the time interval $\tau$ is not an arbitrary time step, but the characteristic time for the elementary physical processes in the model: failure and healing of fibers and the subsequent stress relaxation. 

In the numerical simulation described in the next section, we adopt the unit system in which $\tau=1$. The numerical values for the two rate constants ($k_1$, $\omega$) are thus expressed in terms of $1/\tau$. The unit of stress is chosen to be $\sqrt{G k_B T/\Omega}$, and the unit of energy is $k_B T$. Namely, we set $k_B T=0.1$ and $G k_BT/\Omega =0.1$. Among several parameters that affect the dynamics of the present model, we focus the initial stress $\sigma_0$ and the two rate constants: $k_1$ and $\omega$.


\section{Numerical Results}
In this study, the model is investigated numerically. The number of fibers $N$ is set to be $10^5$,  the ensemble average is taken over $10^4$ samples. The disorder strength, the activation energy and applied stress are kept fixed at $0.5$, $0.1$ and $0.1$ respectively. 

\begin{figure}[ht]
\centering
\includegraphics[width=8.5cm, keepaspectratio]{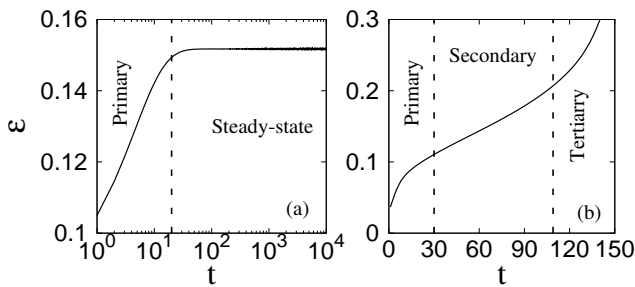} 
\caption{Typical time evolution of strain for (a) larger healing rate ($\omega=0.01$) and (b) smaller healing rate ($\omega=0.001$). Other parameters are kept at $k_BT=0.1$, $\sigma=0.1$, and $\delta=0.5$ for both the panels.}
\label{Strain}
\end{figure}

The time evolution of the system is inspected via strain and strain rate, which are observable in most experiments. 
Since the total load on the system is kept constant during the simulation, the strain at time $t$ is $\epsilon_t=\sigma_0/\phi_t$, where $\sigma_0$ is the initial stress. Since the fraction of intact fibers $\phi_t$ is a decreasing function of time, the strain $\epsilon_t$ is an increasing function of time.

\subsection{Fate of the system}\label{secIIIA}
The dynamic behavior of the system is determined by the competition between the two elementary processes: healing and failure.
As shown in Fig. \ref{Strain}(a), if the healing process is sufficiently effective (i.e., the healing rate is dominant over the rupture rate), the strain increases with time initially and then saturates at a particular value. This is regarded as the steady state and the system does not fail in this case.

Since the strain in this model remains constant at steady states, one may argue whether the model can represent a sheared system, in which strain grows indefinitely with time. We would like to point out that the strain in this model represents the {\it elastic strain}, which is proportional to the stress. Thus, the elastic strain is also finite. It is the {\it inelastic strain} that grows indefinitely with time, and the total strain is the sum of elastic strain and inelastic strain. The inelastic strain is not explicitly modeled here, as it does not involve stress. 

If the healing process is not sufficiently effective, the system fails eventually. Time evolution of the strain rate resembles creep rupture, exhibiting three stages as shown in Fig. \ref{Strain}(b).

\begin{figure}[ht]
\centering
\includegraphics[width=8cm, keepaspectratio]{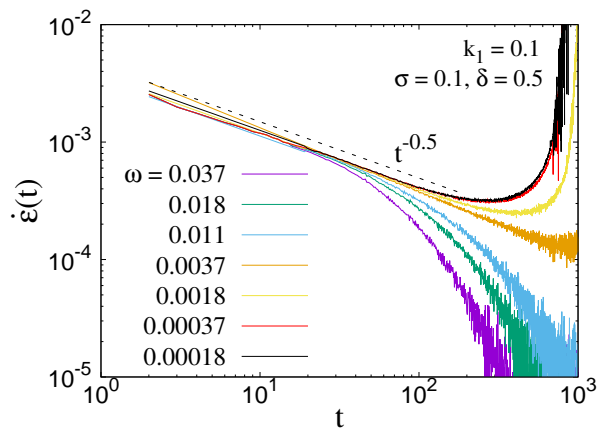} 
\caption{Variation of the strain rate generated in the bundle with increasing time during the course of failure. $k_1$, $k_BT$, $\sigma$, $\delta$ are kept fixed at 0.1, 0.1, 0.1 and 0.5 respectively. $\omega$, on the other hand, is varied continuously. Close to a critical value $\omega_c (\approx 0.0036)$, we observe $\dot{\epsilon}(t) \sim t^{-p}$ where $p=0.5$. For $\omega<\omega_c$, the bundle breaks completely while for $\omega>\omega_c$, it reaches a steady state.}
\label{Fig_Strainrate}
\end{figure}

This qualitative difference in the dynamic behavior is more apparent if the healing rate $\omega$ is varied with the rate of rupture $k_1$ fixed. This is shown in Fig. \ref{Fig_Strainrate}. For smaller $\omega$, healing rate is insufficient to stabilize the system, and the system eventually fails after the acceleration of deformation. For larger $\omega$, the healing rate is dominant over the rupture rate. In this case, the strain rate tends to zero quickly. Namely, deformation stops and the system reaches the steady state. At intermediate values of $\omega$, the strain rate appears to decay as a power law, which is eventually cut off at a certain time scale depending on the healing rate $\omega$. The strain rate vanishes quickly for larger $\omega$, or increases toward failure for smaller $\omega$.

\begin{figure}[ht]
\centering
\includegraphics[width=8cm, keepaspectratio]{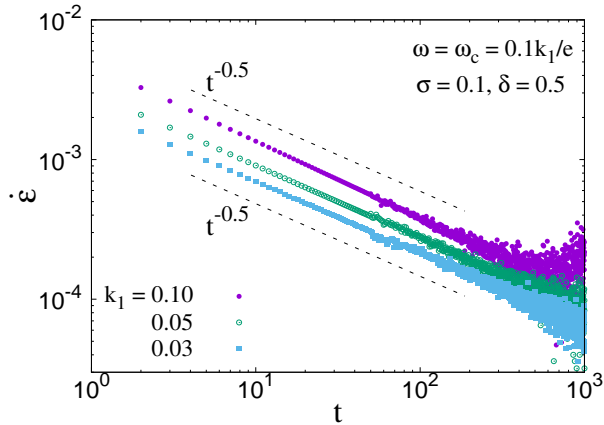} 
\caption{Behavior of the strain rate at critical point: $\omega = \omega_c$. For $k_1=0.1,$ 0.05 and 0.03, the values of $\omega_c$ will close to 0.0036, 0.0018 and 0.0011 respectively. We observe $\dot{\epsilon}(t) \sim t^{-0.5}$ independent of the value of $k_1$. $k_BT$, $\sigma$ and $\delta$ is kept constant at 0.1, 0.1 and 0.5.}
\label{Fig_Critical}
\end{figure}

There seems to exist a critical point ($\omega \rightarrow \omega_c$), at which the strain rate decays in a power-law manner for a long time. In Figure \ref{Fig_Critical}, the power law behavior continues up to three orders of magnitude in time for some values of $k_1$. The strain rate is described as $\dot\epsilon\propto t^{-p}$, and the exponent $p\simeq 0.5$ in the parameter range investigated here.


\subsection{Critical Behavior} 
The strain rate eventually diverges (breakdown) or vanishes (steady state) depending on the healing rate $\omega$.
In Fig. \ref{Fig_Strainrate}, each curve corresponds a different value of $\omega$.

\begin{figure}[ht]
\centering
\includegraphics[width=8cm, keepaspectratio]{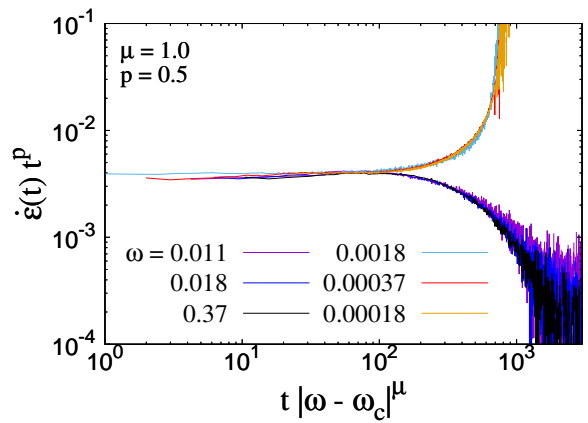} 
\caption{Collapse of strain rate behaviors at various values of $\omega$ with $\mu=1.0$ and $p=0.5$. This implies the validity of Eq. (\ref{scaling}). $k_1$ is kept constant at 0.1.}
\label{Fig_Collapse}
\end{figure}

These behaviors of strain rate $\dot\epsilon(t)$ are described in a much simpler manner using a scaling relation: 
\begin{equation}
\label{scaling}
\dot\epsilon(t) \propto t^{-p}f_{\pm} (t|\omega-\omega_{c}|^{\mu}).
\end{equation}
Here $f(\cdot)_{\pm}$ represents the scaling functions below ($\omega<\omega_{c}$) or above ($\omega>\omega_{c}$) the critical state. It thus implies that the strain rate curves at various values of $\omega$ collapse on a few branches. Figure \ref{Fig_Collapse} shows the collapse of the data onto two branches, which are above/below the critical state. The scaling function at the critical state is constant, and the strain rate is expected to decay as $t^{-p}$. This is shown as Fig. \ref{Fig_Critical}.

\begin{figure}[ht]
\centering
\includegraphics[width=8cm, keepaspectratio]{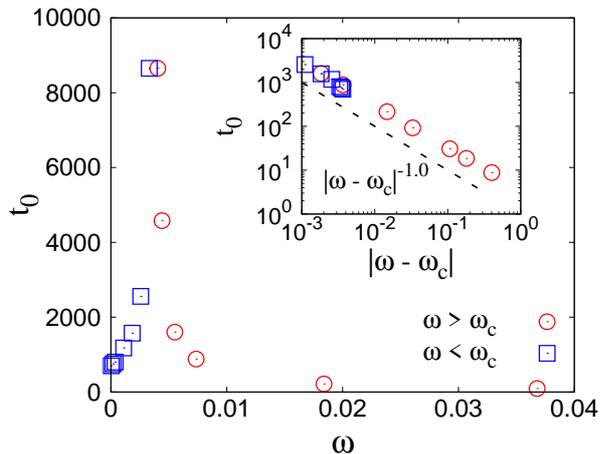} 
\caption{Behavior of the characteristic time $t_0$ associated with the time evolution of the strain rate. $t_0$ is estimated from Eq. \ref{Omori} in the limit $\omega>\omega_c$. For $\omega<\omega_c$, $t_0$ is the time required for the model to fail. $t_0$ diverges around $\omega_c$ in a scale free manner: $t_0 \sim |\omega-\omega_c|^{-\mu}$, where $\mu=1.0$. This is consistent with Fig. \ref{Fig_Collapse} as well as Eq. \ref{scaling}.}
\label{Fig_time}
\end{figure}

\begin{figure}[ht]
\centering
\includegraphics[width=8cm, keepaspectratio]{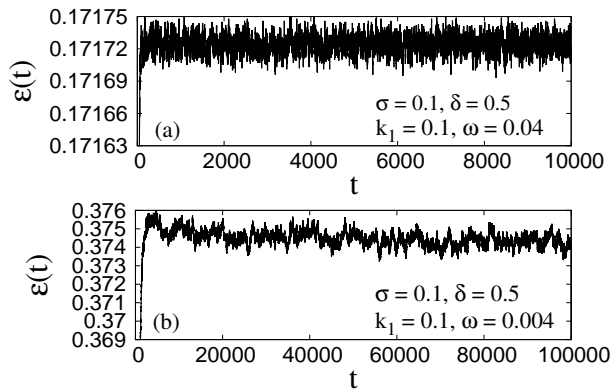}
\caption{Strain $\epsilon(t)$ as a function of time $t$ for the region $\omega>\omega_c$. (a) and (b) shows the results far away from $\omega_c$ and close to $\omega_c$ respectively. We set $k_1=0.1$, $k_BT=0.1$, $\sigma=0.1$ and $\delta=0.5$.}
\label{Fig_strain_fluctuation}
\end{figure}

\begin{figure}[ht]
\centering
\includegraphics[width=8cm, keepaspectratio]{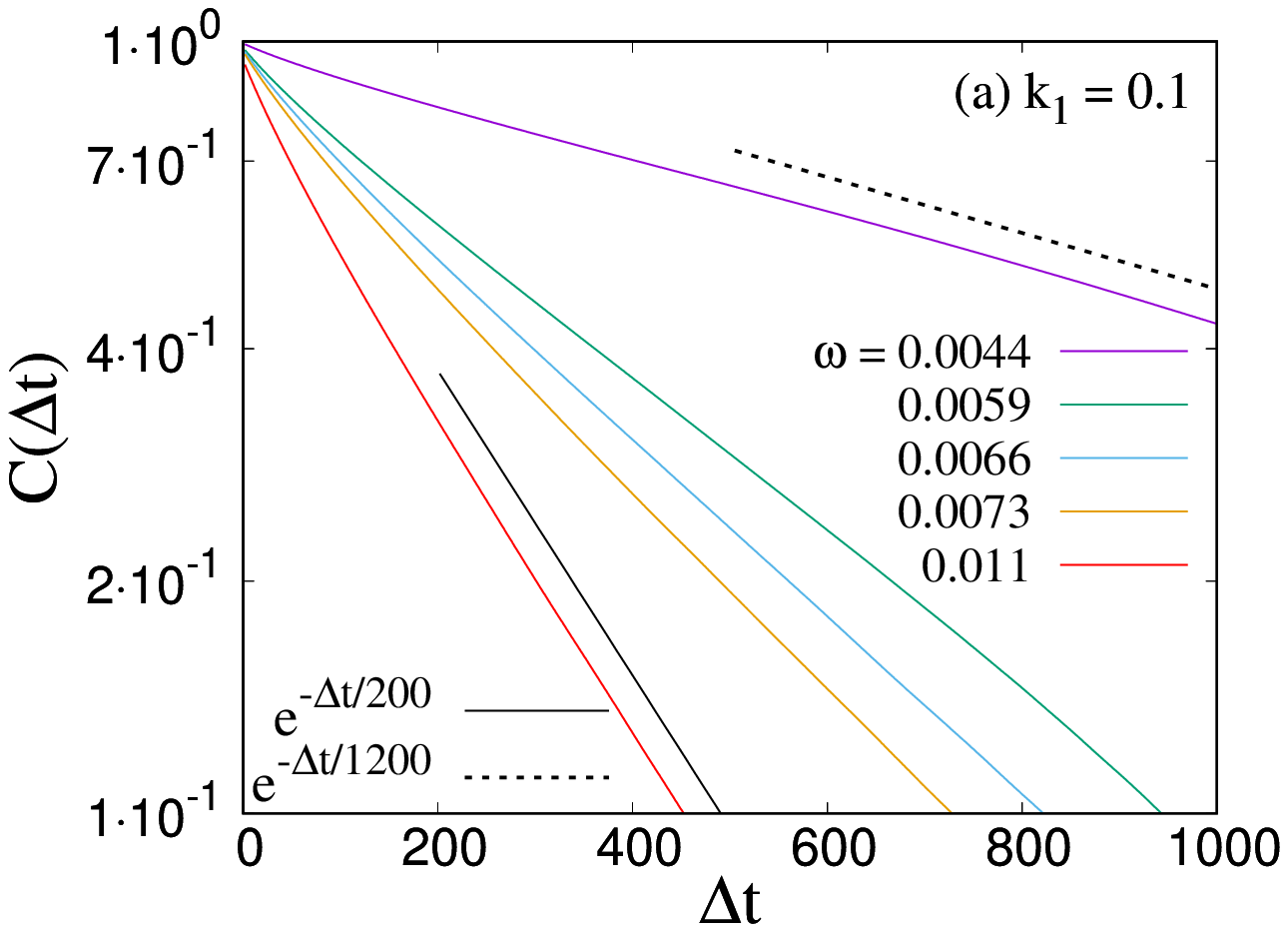}
\includegraphics[width=8cm, keepaspectratio]{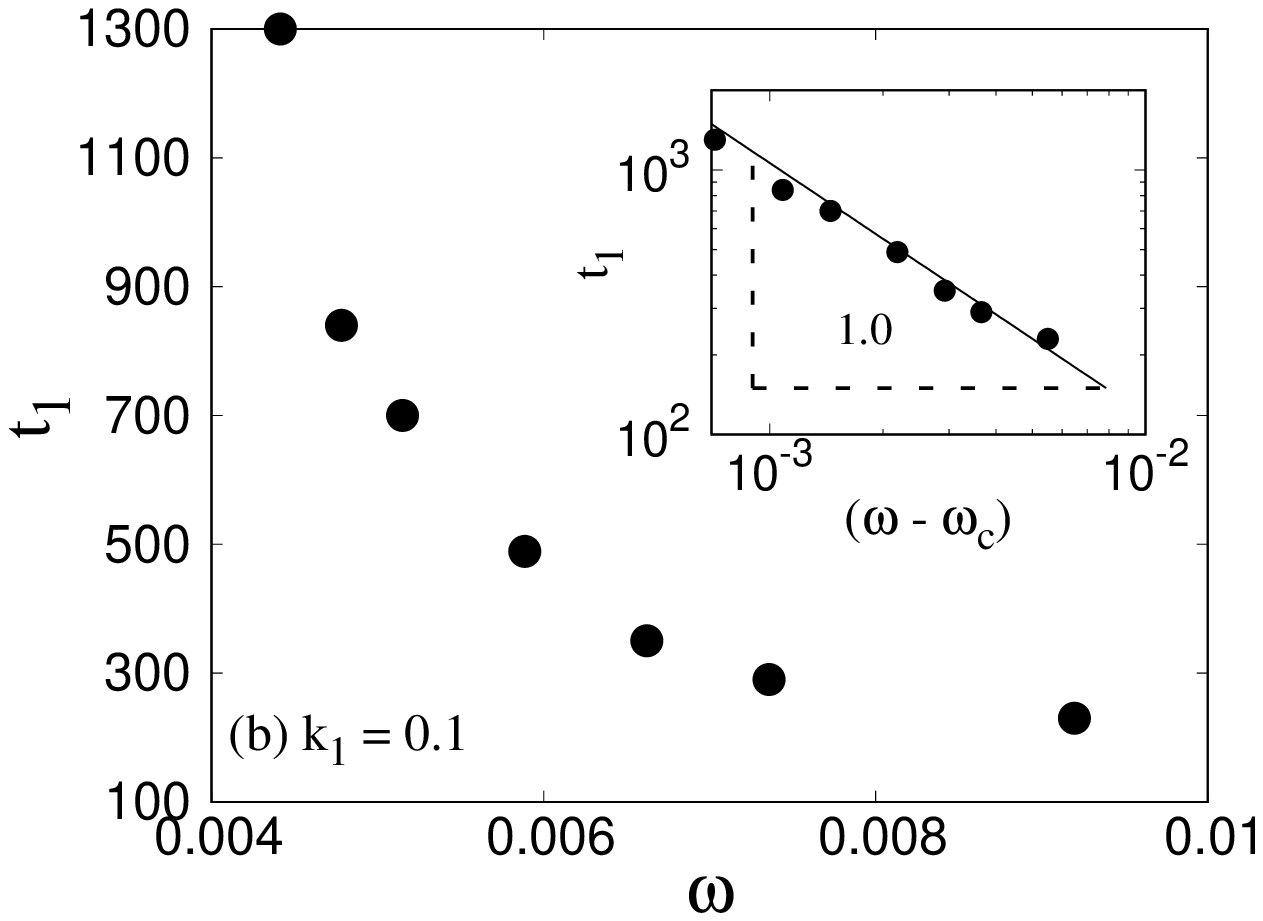}
\caption{(a) Correlation function associated with the $t$ vs $\epsilon(t)$ behavior (see Fig. \ref{Fig_strain_fluctuation}) for $k_1=0.1$, $k_BT=0.1$, $\sigma=0.1$ and $\delta=0.5$. We observe $C(\Delta t) \sim \exp(-\Delta t/t_1)$. The correlation function is longer closer to $\omega_c$. The inset shows the divergence of $t_1$ as $\omega$ approaches $\omega_c$: $t_1 \sim (\omega-\omega_c)^{-\xi}$ with $\xi=1.0$ which is the same value as $\mu$.}
\label{Fig_Correlation_function}
\end{figure}

In addition, the scaling relation (\ref{scaling}) implies the existence of characteristic time that behaves as $|\omega-\omega_{c}|^{-\mu}$. This can be directly confirmed by calculating characteristic time from time series.
For $\omega$ larger than the critical value $\omega_c$, relaxation time to steady state can be defined by fitting the strain rate with the following equation.
\begin{align}\label{Omori}
\dot{\epsilon} \sim (t+c)^{-p}e^{-t/t_0},
\end{align}
where $c$ and $t_0$ are time constants. This behavior is similar to the Omori-Utsu law \cite{Omori,Utsu} for the time evolution of aftershocks rate. Since $c$ involves the initial behavior, the exponential cutoff defines the characteristic time $t_0$ for relaxation. Similarly, another characteristic time for smaller $\omega$ $(\omega< \omega_c)$ is defined as the lifetime of the system. This is also denoted by $t_0$, and shown in Fig. \ref{Fig_time} together with $t_0$ estimated by Eq. (\ref{Omori}) for $\omega>\omega_c$. They both increase toward the critical state obeying $|\omega-\omega|^{-1}$ as shown in Fig. \ref{Fig_time}.
This behavior is consistent with the scaling behavior, Eq. (\ref{scaling}).
  
\subsection{Correlation time for fluctuations in strain}   

The system reaches the steady state for $\omega$ larger than $\omega_c$. The strain $\epsilon$ fluctuates around the mean value. In Fig. \ref{Fig_strain_fluctuation}, time series of strain is shown for the initial $10^4$ time steps. The correlation time seems to be longer in panel (b), where $\omega$ is closer to the critical value.

This is quantitatively shown by computing an auto-correlation function, which are shown in Fig. \ref{Fig_Correlation_function}. The function can be calculated from $n$ values of $\epsilon$ where $n$ stands for increasing time steps. Since the system breakdowns at finite time for $\omega<\omega_c$, the correlation function $C(\Delta t)$ can be computed only for $\omega>\omega_c$, where a steady state is realized. $C(\Delta t)$ has the following form for a series $t$ vs $\epsilon(t)$ in the steady state. $n$ is the number of data points considered to calculate $C(\Delta t)$. $n=5\times10^4$ in our case, after the model is well inside the steady state. 
\begin{align}\label{corr1}
C(\Delta t) = \displaystyle\frac{\displaystyle\frac{1}{n}\displaystyle\sum_{t=1}^{n-\Delta t}[\epsilon(t)-\bar{\epsilon}][\epsilon(t+\Delta t)-\bar{\epsilon}]}{\displaystyle\frac{1}{n}\displaystyle\sum_{t=1}^{n}[\epsilon(t)-\bar{\epsilon}]^2}
\end{align} 
where $\bar{\epsilon}$ is given by,
\begin{align}\label{corr1}
\bar{\epsilon}=\displaystyle\frac{1}{n}\displaystyle\sum_{t=1}^{n}\epsilon(t)
\end{align} 
The correlation function is exponential as shown in Fig. \ref{Fig_Correlation_function}(a). We can thus extract the correlation time $t_1$ by fitting the data with $\exp(-\Delta t/t_1)$. The result is shown in Fig. \ref{Fig_Correlation_function}(b), in which the correlation time increases as $\omega$ decreases to the critical value, $\omega_c$. The behavior suggest the power-law increase of correlation time: $t_1 \propto (\omega-\omega_c)^{-\xi}$. Here the exponent $\xi$ is estimated as 1.0 which is the same value as exponent $\mu$. Both the time constants appear to increase toward the critical point with similar exponents.


\section{Discussions} 

We have studied the fiber bundle model at a finite temperature with a healing mechanism. The temperature, in our case, helps both the rupture and the healing process. Rich relaxation dynamics is observed due to the interplay of the above two factors. The model is observed for different applied stress, temperature, and disorder strength. 

In presence of both probabilistic rupture and infinite healing, the model either stabilizes or meets global failure depending on whether the healing rate constant is higher or lower. In the former case, the strain (or strain rate), generated in the model with time, shows a primary creep followed by a steady-state while for the latter case, all three creep stages: primary, secondary and tertiary are observed. In the case where the steady-state is realized, the strain rate shows the Omori-Utsu like behavior with an exponent which is insensitive to the rate of probabilistic failure.

The critical healing rate, which is a function of the rate of probabilistic failure, is observed to divide the strain rate in two branches: towards steady-state or towards global failure, in a well defined scaling rule. As the healing rate approaches its critical value, the correlation time associated with the time evolution of strain diverges in a scale-free manner.  


\section{Acknowledgment} 

This work was partly supported by the Research Council of Norway through its Centres of Excellence funding scheme, project number 262644. TH is supported by Japan Society for the Promotion of Science (JSPS) Grants-in-Aid for Scientific Research (KAKENHI) Grants Nos. 16H06478 and 19H01811.



\end{document}